\documentclass[journal=jctcce,manuscript=article]{achemso}
\setkeys{acs}{articletitle=true}

\usepackage[version=3]{mhchem} 
\usepackage{hyperref}
\usepackage{xcolor}
\usepackage{graphicx}
\usepackage{amsmath}
\usepackage{physics}
\usepackage{bm}
\usepackage{booktabs}
\usepackage{chemformula}
\usepackage{cleveref}
\usepackage{footnote}

\renewcommand{\H}{\mathcal{\hat{H}}}
\newcommand{\Ht}{\mathcal{\tilde{H}}}
\newcommand{\E}{\hat{E}}

\newcommand{\g}{\gamma}
\newcommand{\G}{\Gamma}
\newcommand{\Fi}{F^{I}}
\newcommand{\Fa}{F^{A}}
\newcommand{\Fit}{\tilde{F}^{I}}
\newcommand{\Fat}{\tilde{F}^{A}}
\newcommand{\Qt}{\tilde{Q}}
\author{Tommaso Nottoli}
\affiliation{Dipartimento di Chimica e Chimica Industriale, Universit\`a di Pisa. Via G. Moruzzi 13, I-56124 Pisa, Italy}
\author{J\"urgen Gauss}
\affiliation{Department Chemie, Johannes Gutenberg-Universit\"at Mainz, Duesbergweg 10-14, D-55128 Mainz, Germany}
\author{Filippo Lipparini}
\email{filippo.lipparini@unipi.it}
\affiliation{Dipartimento di Chimica e Chimica Industriale, Universit\`a di Pisa. Via G. Moruzzi 13, I-56124 Pisa, Italy}

\title
  {A Second-Order CASSCF Algorithm with the Cholesky Decomposition of the Two-Electron Integrals}

\abbreviations{CASSCF}

\begin{document}

\begin{tocentry}
\includegraphics{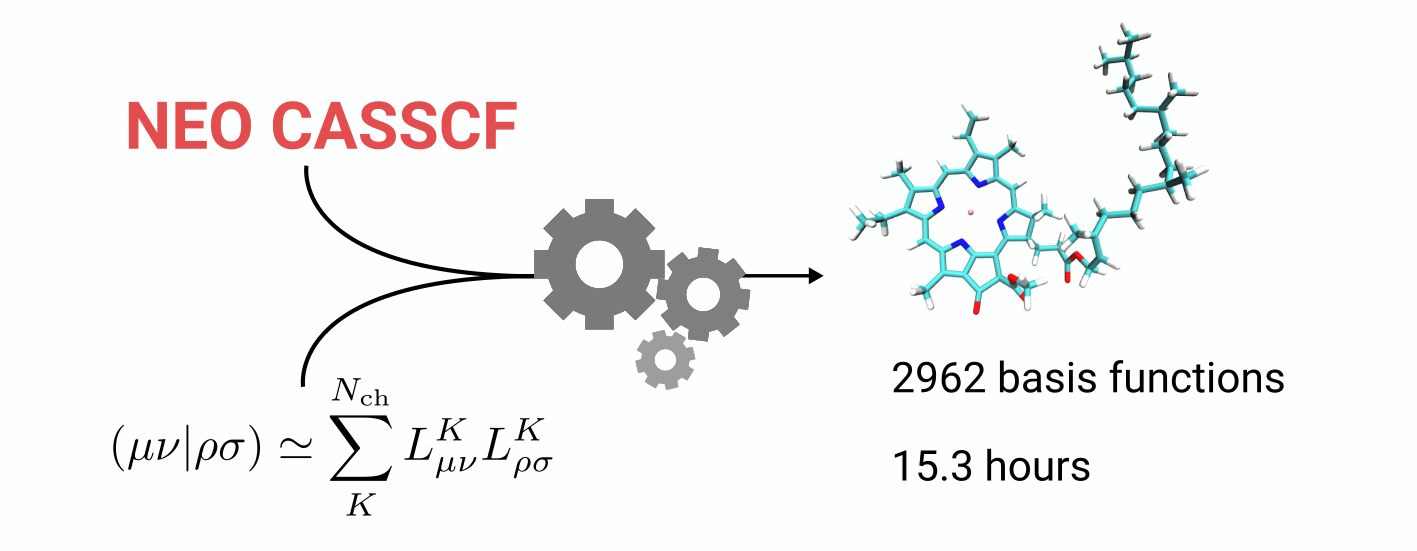}
\end{tocentry}

\begin{abstract}
In this contribution, we present the implementation of a second-order CASSCF algorithm in conjunction with the Cholesky decomposition of the two-electron repulsion integrals. The algorithm, called Norm-Extended Optimization, guarantees convergence of the optimization, but it involves the full Hessian of the wavefunction and is therefore computationally expensive. Coupling the second-order procedure with the Cholesky decomposition leads to a significant reduction in the computational cost, reduced memory requirements, and an improved parallel performance. As a result, CASSCF calculations of larger molecular systems become possible as a routine task. The performance of the new implementation is illustrated by means of benchmark calculations on molecules of increasing size, with up to about 3000 basis functions and 14 active orbitals.  
\end{abstract}

\section{Introduction}
The complete active space--self-consistent field (CASSCF) method\cite{Werner1987,Shepard1987,Roos1987} is a powerful tool to achieve a qualitatively correct description of strongly correlated systems. Thanks to its intrinsic multireference nature, it can be used to compute the structure and molecular properties of a large manifold of interesting systems that are poorly described with standard single-reference methods. These include many open-shell systems, molecules with stretched bonds, and therefore reactivity, excited states and others. It can also provide a starting point for subsequent high-level correlated treatments, such as internally-contracted multireference configuration interaction\cite{Knowles1988,Werner1988} (CI) and coupled cluster\cite{Jeziorski1981,Banerjee1981,Aoto2016,Hanauer2011,Kohn2013,Evangelista2011,Evangelista2012,Lipparini2017}; multireference perturbation theory such as CASPT2\cite{Andersson1990,Andersson1992}, and NEVPT2\cite{Angeli2001,Angeli2001a,Angeli2002}; or even quantum Monte Carlo methods\cite{QMCRev,QMC2}. 
Unfortunately, the method suffers of three major complications that restrict its applicability. First, it is not a black box method, as it requires the user to select the active space for the calculation. While there are a few strategies to aid the selection\cite{Stein2016,Sayfutyarova2017,Toth2020}, achieving good results relies still on the user's chemical intuition and understanding of the system. Second, the CASSCF wavefunction's optimization problem is notoriously hard to converge. Third, the method is computationally very demanding.

The computational cost of a CASSCF calculation stems from two concurring factors. The most prominent one is that the method requires to solve a full CI (FCI) problem in the active space. Due to the combinatorial scaling of FCI, the investigation of large active spaces is not possible using standard direct CI techniques. Approximations to the FCI wavefunction can be used to overcome this otherwise overwhelming barrier, the most common example being the use of a density-matrix renormalization group\cite{Chan2011} (DMRG). However, many interesting systems can be successfully described with a relatively small active space (up to 12-14 electrons in as many orbitals). If a careful choice of the active space that allows to capture the static correlation of the wavefunction with a limited number of active orbitals is possible, the cost of the CI part is either negligible (for active spaces with less than 10 orbitals) or manageable with traditional implementations. In such cases, the cost of the calculation is dominated by the operations involving the manipulation of the electron repulsion integrals (ERIs). 

Convergence problems can be mitigated, if not completely solved, by using an optimization algorithm with guaranteed convergence to the closest local minimum. Methods based on a restricted step second-order optimization offer such a guarantee and are, therefore, a very attractive option. However, as they involve the evaluation of the energy Hessian with respect to the variational parameters, i.e., orbital rotations and CI expansion coefficients, they are in general more expensive than their first-order counterparts and require cumbersome and involved implementations. Nevertheless, second-order CASSCF implementations have been successfully achieved and are based on two main algorithm. The first algorithm, originally proposed by Werner and Meyer\cite{Werner1980} and further refined by Werner, Knowles and others\cite{Werner1985,Kreplin2019}, is based on the definition of a model energy function which is infinite order in orbital rotations and that is optimized. The coupling between CI and orbital optimization is introduced up to the second-order, ensuring thus quadratic convergence. This algorithm shows excellent convergence properties and overall performances. A similar strategy has been followed by Sun et al.\cite{Sun17} and the resulting algorithm, which is based on an integral-direct implementation and can use DMRG as a CASSCF solver, exhibits impressive performances. 
A second choice is to use a more traditional trust-region second-order method, such as the Levenberg-Marquardt method\cite{Fletcher1999}. Augmented with an adaptive choice of the trust radius, as proposed by Fletcher (we refer to the global strategy as FLM), it is possible to prove that the FLM method always converges to the closest local minimum and that the rate of convergence is quadratic. A very efficient implementation of the FLM method, known as the Norm-Extended Optimization (NEO) algorithm, has been proposed by Jensen and coworkers\cite{Jensen1983,Jensen1986}. In this contribution, we follow the latter strategy, which we have previously implemented in the CFOUR\cite{cfour,Matthews2020} suite of programs. 

A second-order CASSCF implementation requires to work with ERIs transformed in the molecular orbital (MO) basis with at least two indices spanning the full rank of MOs. The transformation of the ERIs from the atomic orbitals (AO) to the MO basis is expensive, requiring $\mathcal{O}(MN^4)$ floating point operations, where $M$ is the number of internal and active orbitals and $N$ the number of basis functions. Furthermore, it is not easily implemented in an efficient way. This is due to the fact that the ERIs matrix is usually too large to fit in memory, especially in the MO basis, which implies that the transformation involves slow disk I/O. Furthermore, the AO ERIs are computed (and stored) in an order that depends on the shell structure of the basis set for the specific system. As a consequence, the integrals are read (or recomputed, in integral direct implementations) in a system-dependent order, which makes the use of efficient BLAS routines\cite{Lawson1979,Dongarra1990} and, more in general, vectorization, particularly challenging.

To address the computational cost involved with the manipulation of the ERIs, it is possible to adopt a low-rank approximation of the ERIs, such as density fitting\cite{Whitten1973,Dunlap79,Vahtras93,Feyereisen1993,Eichkorn95,Weigend97,Weigend2002,Sierka03,Sodt2006} (DF) or Cholesky Decomposition\cite{Beebe1977,Roeggen1986,Koch2003,Roeggen2008,Aquilante2011,Weigend2009} (CD). Both techniques have been successfully applied in many contexts of quantum chemistry\cite{Werner2003,Manby2003,Polly2004,Bostrom2013,Bozkaya2014,Bozkaya2014a,Bozkaya2016,Bozkaya2016a,Bozkaya2017}, including CASSCF\cite{Aquilante2008,Delcey2014,Reynolds2018}. 
The CD technique is particularly attractive, as it allows a rigorous, a priori control of the approximation error. Furthermore, it offers a compact representation of the ERIs that is well suited for vector, efficient implementations, as the Cholesky-decomposed ERIs can be often kept in memory with standard computer hardware and are easily manipulated using highly optimized level 3 BLAS routines. Furthermore, all the ERIs manipulation can be written as the sum of independent operations on a given Cholesky vector and are therefore very easy to parallelize.

In this contribution, we present an implementation of NEO CASSCF in the CFOUR suite of programs\cite{cfour,Matthews2020} based on the CD of the ERIs. The implementation is tested on several molecular systems of increasing size, for active spaces that go from small (CAS(6,6)) to large (CAS(14,14)) and using up to about 3000 basis functions. The paper is organized as follows. In Section~\ref{sec:theory}, the derivation of the NEO CASSCF method is reviewed. The implementation of the algorithm is discussed in Section~\ref{sec:imple} with a special focus on the Cholesky implementation. In Section~\ref{sec:bench}, benchmark calculations are presented for the purpose of showing the performance of the algorithm in the optimization of medium-to-large systems. Finally, concluding remarks and some perspectives on future developments are given is Section~\ref{sec:concl}.

\section{Norm Extended Optimization CASSCF}\label{sec:theory}
In this section, we recapitulate the main aspects of NEO CASSCF. First, the parametrization of the wavefunction is discussed in Section~\ref{sec:cas-wvf}. Then, the NEO algorithm is briefly summarized in Section~\ref{sec:cas-opt}. 
Further details regarding the optimization algorithm can be found in Ref.~\citenum{Jensen1983} or in a previous paper by two of us \cite{Lipparini2016}.

\subsection{Parametrization of the CASSCF wavefunction\label{sec:cas-wvf}}
The starting point for the following discussion is given by a set of molecular orbitals (MOs) $\{\varphi_{p}\}_{p=1}^{N_{b}}$, where $N_{b}$ is the number of basis functions. 
In CASSCF, the MOs are subdivided into three classes according to their allowed occupation number in a Slater determinant\textemdash namely \textit{internal}, which are always doubly occupied; \textit{active}, which are subjected to no restriction; and \textit{external}, which are always empty.
To distinguish an orbital among such classes, the following labels are used: $i,j,k$ refer to inactive, $u,v,x$ to active, $a,b,c$ to external, and $p,q,r$ to generic orbitals. Indices that run over the determinantal space are labelled with capital letters $I,J$.
\par
A convenient parametrization for the wavefunction, first proposed by Jensen and J\o rgensen\cite{Jensen1983}, is
\begin{equation}\label{eq:wvf}
 \ket{\Psi} = e^{-\hat{\kappa}}\frac{\ket{0}+\hat{P}\ket{\bm{c}}}{\norm{\ket{0}+\hat{P}\ket{\bm{c}}}}.
\end{equation}
Here, $\ket{0}=\sum_{I}^{N_{\text{det}}}c^{(0)}_{I}\ket{\Phi^{(0)}_{I}}$ is the current approximation to the wavefunction, or current expansion point (CEP). 
$\ket{\bm{c}}$ is the correction vector that collects the CI variational parameters $c_{I}$,
\begin{equation}
 \ket{\bm{c}} = \sum_{I}^{N_{\text{det}}}c_{I}\ket{\Phi_{I}},
\end{equation}
and $\hat{P}=1-\ket{0}\bra{0}$ is the operator that projects $\ket{\bm{c}}$ in the orthogonal complement of $\ket{0}$ thus keeping any redundant vector parallel to the CEP. 
\par
Orbitals variations are described through a unitary transformation, $\bm{\bar{\varphi}}=\bm{\varphi}\bm{U}$, 
that is conveniently parametrized by using an exponential map
\begin{equation}
 U=e^{-\kappa}; \quad \hat{\kappa} = \sum_{p>q}\kappa_{pq}\left(\hat{E}_{pq}-\hat{E}_{qp}\right) = \sum_{p>q}\kappa_{pq}\hat{E}^{-}_{pq}
\end{equation}
where $\E_{pq}=\sum_{\sigma}\hat{a}^{\dagger}_{p\sigma}\hat{a}_{q\sigma}$ is the spin-traced singlet excitation operator. 
The variational parameters are given by the elements of the anti-symmetric matrix, $\bm{\kappa}$.
Since only rotations between different orbitals classes produce a variation in the energy, the expression for $\hat{\kappa}$ can be simplified as follow
\begin{equation}
 \hat{\kappa} = \sum_{i}^{N_{\text{int}}}\sum_{u}^{N_{\text{act}}}\kappa_{iu}\E^{-}_{iu} +
                \sum_{i}^{N_{\text{int}}}\sum_{a}^{N_{\text{ext}}}\kappa_{ia}\E^{-}_{ia} +
                \sum_{a}^{N_{\text{ext}}}\sum_{u}^{N_{\text{act}}}\kappa_{au}\E^{-}_{au}.
\end{equation}
Hence, $\bm{\kappa}$ is considered as a vector whose dimension is given by all non-redundant orbitals rotations, \textit{i.e.} $N_{\text{rot}} = N_{\text{int}}N_{\text{act}} + N_{\text{int}}N_{\text{ext}} + N_{\text{act}}N_{\text{ext}}$.

\subsection{Optimization of the CASSCF wavefunction}\label{sec:cas-opt}
Equation \ref{eq:wvf} is used to define a variational expression for the electronic energy that reads
\begin{equation}\label{eq:var-ene}
 \mathcal{E}(\bm{\kappa},\bm{c}) = \frac{\left(\bra{0}+\bra{\bm{c}}\hat{P}\right)
                                   e^{\hat{\kappa}}\H e^{-\hat{\kappa}}
                                   \left(\ket{0}+\hat{P}\ket{\bm{c}}\right)}
                                   {1 + \bra{\bm{c}}\hat{P}\ket{\bm{c}}}.
\end{equation}
In equation \ref{eq:var-ene}, $\H$ is the non-relativistic Hamiltonian operator written in second quantization
\begin{equation}
 \H = \sum_{pq}h_{pq}\E_{pq} + \frac{1}{2}\sum_{pqrs}(pq|rs)\hat{e}_{pqrs} + E_{\text{nuc}}
\end{equation}
where
\begin{equation}
 \hat{e}_{pqrs} = \E_{pq}\E_{rs} - \delta_{rq}\E_{ps},
\end{equation}
$h_{pq}$ are one-electron integrals, $(pq|rs)$ are two-electron integrals written in Mulliken's notation, and $E_{\rm nuc}$ is the nuclear repulsion term.
A second-order algorithm can be developed by defining a quadratic model for the energy; therefore, we expand equation~\ref{eq:var-ene} in power series up to second order. To this end, it is useful to define a generic parameter point, $\bm{x} = (\bm{c}, \bm{\kappa})$, and the reference one, $\bm{x}_{0} = (\bm{c}^{(0)}, \bm{0})$ such that
\begin{equation}\label{eq:ene-taylor}
 \mathcal{E}(\bm{x})\approx \mathcal{Q}(\bm{x}) \doteq E_{0} + \bm{g}^{\dagger}(\bm{x}-\bm{x}_{0}) 
 + \frac{1}{2}(\bm{x}-\bm{x}_{0})^{\dagger}\bm{G}(\bm{x}-\bm{x}_{0}).
\end{equation}
In equation \ref{eq:ene-taylor}, $E_{0}$ is the reference energy, that is $\bra{0}\H\ket{0}$, while $\bm{g}$ and $\bm{G}$ are respectively the electronic
gradient and Hessian evaluated at the CEP. Analytical expression for such quantities can be obtained by direct differentiation of equation~\ref{eq:ene-taylor} and by exploiting the Baker-Campbell-Hausdorff (BCH) formula.
The gradient is given as
\begin{align}
 & g^{c}_{I}  = \pdv{\mathcal{E}}{c_{I}} = 2\bra{\Phi_{I}}\hat{P}\H\ket{0}, \label{eq:ci-grad-general}\\
 & g^{o}_{pq} = \pdv{\mathcal{E}}{\kappa_{pq}} = \bra{0}[\E^{-}_{pq},\H]\ket{0},\label{eq:orb-grad-general}
\end{align}
and the Hessian
\begin{equation}
 \begin{split}
  & G^{cc}_{I,J}  = \pdv{\mathcal{E}}{c_{I}}{c_{J}} = 2\bra{\Phi_{I}}\hat{P}(\H-E_{0})\ket{\Phi_{J}}, \\
  & G^{co}_{I,pq} = \pdv{\mathcal{E}}{c_{I}}{\kappa_{pq}} = 2\bra{\Phi_{I}}\hat{P}[\E^{-}_{pq},\H]\ket{0},\\
  & G^{oo}_{pq,rs}= \pdv{\mathcal{E}}{\kappa_{pq}}{\kappa_{rs}} = 
                    \frac{1}{2}(1+\hat{P}_{pq,rs})\bra{0}\hat{P}[\E^{-}_{pq},[\E^{-}_{rs},\H]]\ket{0}.
 \end{split}
\end{equation}
The minimization of the quadratic model directly leads to the Newton-Raphson (NR) equations. However, the radius of convergence of NR is small, and the Hessian can be non positive-definite at the beginning of the optimization leading to incorrect search directions. To overcome this issue, a more robust strategy consists in using a trust-region optimization algorithm, \textit{e.g.} the Levenberg-Marquardt (LM) method\cite{Fletcher1999}, where the minimization is performed in a reduced domain such that the Hessian has the correct signature. The LM equations can be seen as diagonally shifted NR ones, where the shifting parameter controls the step length to be within a predefined trust radius $R_{t}$.
The Norm-Extended Optimization (NEO) algorithm \cite{Jensen1983,Jensen1987} is an elegant way to recast the LM minimization problem into an eigenvalue-eigenvector one
\begin{equation}
 \bm{L}(\alpha)\bm{y} = \lambda\bm{y}
\end{equation}
where $\bm{L}(\alpha)$ is the gradient-scaled augmented Hessian matrix
\begin{equation}
 \bm{L}(\alpha) = \bm{G} + \alpha\left(\bm{x}_{0}\bm{g}^{\dagger} + \bm{g}^{\dagger}\bm{x}_{0}\right).
\end{equation}
It can be shown that for ground-state optimization the optimal direction is given by the first eigenvector of $\bm{L}(\alpha)$. Once $\bm{y}$ is given, the NEO step can be computed as
\begin{equation}
 \bm{x}(\alpha) = \bm{x}_{0} + \frac{1}{\alpha}\bm{P}\bm{y}(\alpha).
\end{equation}
Here, $\bm{P}$ is the matrix representation of the projector operator $\hat{P}$.
The step length is controlled by the parameter $\alpha$ and can be obtained by solving the equation
\begin{equation}
 \norm{\bm{x}(\alpha)-\bm{x}_{0}} = R_{t}.
\end{equation}
Eventually, the trust radius is changed adaptively during the optimization procedure according to
Fletcher's algorithm \cite{Fletcher1999}. If the energy increases, the step is discarded, and the trust radius is decreased.
Otherwise, $R_{t}$ is either increased or left untouched based on the value of the ratio between the predicted variation of the energy and the actual one.
The combined strategy\textemdash NEO plus Fletcher's update\textemdash leads to an algorithm that always converge to the closest local minimum for well behaved wavefunctions.

\section{Implementation}\label{sec:imple}
In this section, the implementation of the NEO algorithm within the CFOUR \cite{cfour,Matthews2020} suite of programs is discussed.
In Section~\ref{sec:direct-eqs}, we present working expressions for the gradient and for the linear transformations that describe the action of the augmented Hessian matrix on a trial vector.
In Section~\ref{sec:cd-eqs}, the Cholesky Decomposition of the two-electron integrals is introduced. 
Details regarding a cost-effective implementation that exploits the Cholesky vectors are given for a specific example.

\subsection{Direct NEO equations}\label{sec:direct-eqs}
The NEO algorithm can be thought as a two-level procedure. In the first level\textemdash the macro-iterations\textemdash the parameter hyper-surface is scanned by updating the CEP and the MOs. 
In the second level\textemdash the micro-iterations\textemdash a specific NEO eigenvalue-eigenvector problem is iteratively solved with the intention of getting the optimal step direction.
At each macro-iteration an atomic orbitals (AO) to MO transformation is performed. Then, the orbital and CI gradient are assembled, and the electronic energy is calculated.
The CASSCF energy can be written as
\begin{equation}
 E_{0} = \sum_{uv}\g_{uv}\Fi_{uv} + \frac{1}{2}\sum_{uvxy}\G_{uvxy}(uv|xy) + E_{i} + E_{\text{nuc}} 
\end{equation}
where $\g_{uv}$ and $\G_{uvxy}$ are the one- and two-body reduced density matrices respectively, which can be computed as the expectation value of the excitation operators
\begin{equation}
 \g_{uv} = \bra{0}\E_{uv}\ket{0}, \quad \G_{uvxy} = \bra{0}\hat{e}_{uvxy}\ket{0}.
\end{equation}
$\Fi_{pq}$ are the elements of the inactive Fock matrix
\begin{equation}\label{eq:fi}
 \Fi_{pq} = h_{pq} + \sum_{i}\left[2(pq|ii)-(pi|qi)\right],
\end{equation}
and $E_{i}$ is the energy contribution that stems from the inactive electrons and is called inactive energy
\begin{equation}
 E_{i} = \sum_{i}\left(h_{ii} + \Fi_{ii}\right).
\end{equation}
Manipulation of equation \ref{eq:orb-grad-general} leads to an anti-symmetric expression for the orbital gradient
\begin{equation}\label{eq:orb-grad}
 g_{pq} = 2(F_{pq} - F_{qp}).
\end{equation}
In equation \ref{eq:orb-grad} we have introduced the generalized Fock matrix, whose elements can be written in terms of the inactive Fock matrix, active Fock matrix, and Q matrix.
The last two are defined as follows:
\begin{align}
 &\Fa_{pq} = \sum_{uv}\g_{uv}\left[(pq|uv) - \frac{1}{2}(pu|qv)\right],\label{eq:fa}\\
 &Q_{up} = \sum_{vxy}\G_{uvxy}(pv|xy)\label{eq:q}.
\end{align}
As equation \ref{eq:ci-grad-general} states, the CI gradient can be evaluated as the action of the Hamiltonian operator on $\ket{0}$
\begin{equation}\label{eq:ci-grad}
 g_{I} = 2\bra{\Phi_{I}}\sum_{uv}\Fi_{uv}\E_{uv} + \frac{1}{2}\sum_{uvxy}(uv|xy)\hat{e}_{uvxy}\ket{0} + c^{(0)}_{I}(E_{0}-E_{i})
\end{equation}
where the last term is a vector parallel to the CEP that stems from the presence of the projector operator
in the wavefunction definition.
\par
The iterative solution of the NEO eigenvalue-eigenvector problem (micro-iterations) requires 
setting up expressions for the matrix-vector product between the augmented Hessian and a trial 
vector
\begin{equation}
 \begin{pmatrix}
  \bm{L}^{cc} & \bm{L}^{co} \\
  \bm{L}^{oc} & \bm{L}^{oo}
 \end{pmatrix}
 \begin{pmatrix}
  \bm{v}^{c}\\
  \bm{v}^{o}
 \end{pmatrix}
 =
 \begin{pmatrix}
  \bm{L}^{cc}\bm{v}^{c} \\
  \bm{L}^{oc}\bm{v}^{c}
 \end{pmatrix}
 +
 \begin{pmatrix}
  \bm{L}^{co}\bm{v}^{o} \\
  \bm{L}^{oo}\bm{v}^{o}
 \end{pmatrix}.
\end{equation}
The present implementation makes use of the split-Davidson algorithm \cite{Jensen1987}, where configurations-only, $\bm{v}_{\text{conf}}=(\bm{v}^{c}, \bm{0})$, and orbitals-only, $\bm{v}_{\text{orb}}=(\bm{0}, \bm{v}^{o})$, vectors are added in the Krylov-like subspace. This procedure allows to adaptively add to the subspace either $\bm{v}_{\text{conf}}$ or $\bm{v}_{\text{orb}}$ depending on the part that exhibits the largest residual.
Here we report the expressions for the direct product
\begin{align}
&\sum_{J} L_{I,J}v_{J} = 2\left(\bra{\Phi_{I}}\H\ket{\bm{v}_{c}} - E_{0}v_{I}\right)
                                  + (\alpha-1)\left[c^{(0)}_{I}\sum_{J}v_{J}g_{J} +
                                  g_{I}\sum_{J}v_{J}x_{J}\right],\label{eq:cc-block} \\
&\sum_{I} L_{pq,I}v_{I} = 2g^{T}_{pq} + \left(\alpha
                                   - 2\right)g_{pq}\sum_{I}c^{(0)}_{I}v_{J},\label{eq:oc-block}\\
&\sum_{pq} L_{I,pq}v_{pq} = 2\bra{0}\Ht\ket{\Phi_{I}} + \left(\alpha - 2\right)c^{(0)}_{I}
                                 \sum_{pq}g_{pq}v_{rs},\label{eq:co-block} \\
&\sum_{rs}L_{pq,rs}v_{rs} = \bra{0}\left[\E^{-}_{pq},\Ht\right]\ket{0} 
                                      +\frac{1}{2}\sum_{s}(g_{sp}v_{qs}-g_{sq}v_{ps}),\label{eq:oo-block}
\end{align}
where $\ket{\bm{v}_{c}}=\sum_{I}v_{I}\ket{\Phi_{I}}$, and $\Ht=\left[\hat{\kappa},\H\right]$ is the one-index transformed Hamiltonian operator.
In equation \ref{eq:oc-block}, we have introduced the transition gradient $g^{T}_{pq}$; that is, a gradient computed with symmetrized transition density matrices.
The first term of equation \ref{eq:oo-block} is a gradient-like contribution computed with one-index transformed one- and two-electron integrals. It can be effectively computed by means of the transformed inactive Fock matrix, active Fock matrix, and Q matrix whose expressions are given below
\begin{align}
 &\Fit_{pq} = \sum_{r}\left(\Fi_{pr}v_{qr}-\Fi_{rq}v_{pr}\right)
              +\sum_{r}\sum_{i}\left[4(pq|ir)-(pr|qi)-(pi|qr)\right]v_{ir},\\
 &\Fat_{pq} = \sum_{r}\left(\Fa_{pr}v_{qr}-\Fa_{rq}v_{pr}\right)
              +\sum_{r}\sum_{uv}\g_{uv}\left[2(pq|ur)-\frac{1}{2}(pr|uq)-\frac{1}{2}(pu|qr)\right]v_{ur},\\
 &\Qt_{up}  = \sum_{r}Q_{ur}v_{pr}
              +\sum_{r}\sum_{vxy}\Big\{\G_{uvxy}(pr|xy) + [\G_{uxvy}+\G_{uxyv}](px|ry)\Big\}v_{vr}. \label{eq:qt}
\end{align}
Explicit expressions for equation \ref{eq:oo-block} are given in the Supporting Information.
\par
\begin{figure}
 \centering
 \includegraphics[width=0.9\textwidth]{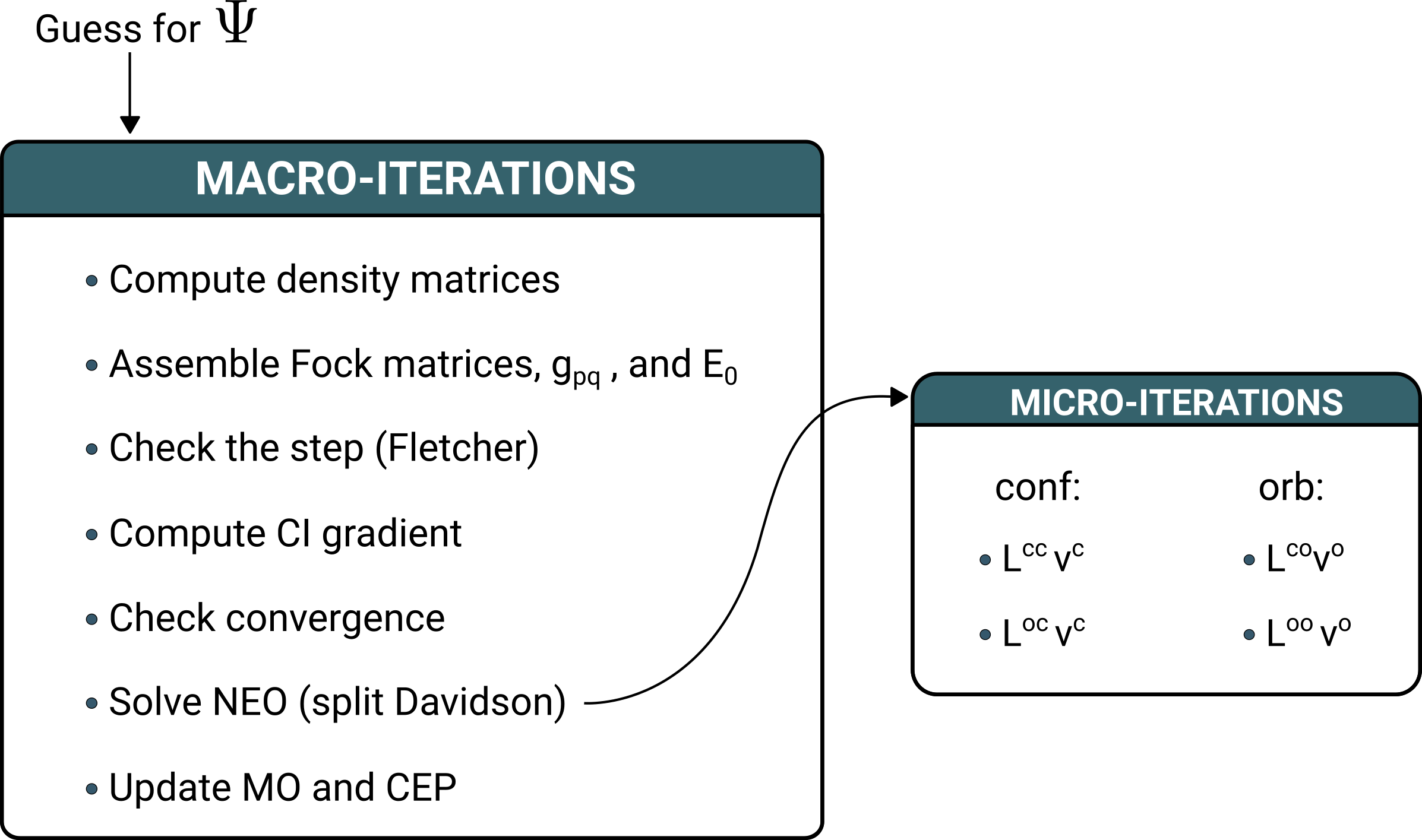}
 \caption{Macro/micro iterations scheme for the NEO algorithm}
 \label{fig:neo-alg}
\end{figure}

The transformed matrices have to be computed at each step of the micro-iterations and together with the AO to MO transformation constitute the bottleneck of the algorithm when the chosen active space is small.
A summary of the NEO algorithm is given in Figure \ref{fig:neo-alg}.

\subsection{NEO equations with Cholesky vectors}\label{sec:cd-eqs}
The ERI matrix is symmetric and positive semidefinite; therefore, it can be decomposed according to the Cholesky Decomposition (CD)
\begin{equation}\label{eq:ch-dec}
 (pq|rs) \simeq \sum_{K}^{N_{\text{ch}}}L^{K}_{pq}L^{K}_{rs}.
\end{equation}
We compute the CD of the integrals using the partial pivoting algorithm proposed by Koch et al.\cite{Koch2003}, which has been implemented inside the Mainz integral package\cite{mint} (MINT) in CFOUR\cite{cfour,Matthews2020}. The procedure stops whenever the residual of the diagonal is below a user defined threshold. Using the Cauchy-Schwarz inequality, it can be shown that the error on the reconstructed integrals is always lower equal than the threshold, so it can be controlled systematically. In equation \ref{eq:ch-dec}, $N_{\text{ch}}$ is the number of Cholesky vectors generated; the higher the decomposition threshold the lower the number of Cholesky vectors. 
\par
The Cholesky representation of the integrals has been substituted in all equations, namely the Fock matrices, the transformed Fock matrices, and the active ERI matrix. In order to illustrate the implementation of the evaluation of the aforementioned quantities, we discuss in detail the calculation of the transformed Q-matrix. Implemented expressions for the transformed Fock matrices can be found in the Supporting Information.
Inserting equation \ref{eq:ch-dec} into equation \ref{eq:qt} we get:
\begin{equation}\label{eq:qt-cd}
 \Qt_{up}  = \sum_{r}Q_{ur}v_{pr}
              +\sum_{K}^{N_{\text{ch}}}\sum_{r}\sum_{vxy}
              \Big\{\G_{uvxy}L^{K}_{pr}L^{K}_{xy} + [\G_{uxvy}+\G_{uxyv}]L^{K}_{px}L^{K}_{ry}\Big\}v_{vr}.
\end{equation}
The first term of equation \ref{eq:qt-cd} can be straightforwardly computed from the Q matrix. The second term is evaluated by first assembling, for each Cholesky vector, the intermediate quantities
\begin{equation}
 T^{K}_{uv} = \sum_{xy}\G_{uvxy}L^{K}_{xy}
\end{equation}
and
\begin{equation}\label{eq:int-s}
 S^{K}_{vp} = \sum_{r}v_{vr}L^{K}_{rp}
\end{equation}
such that 
\begin{equation}\label{eq:int-ts}
 \sum_{r}\sum_{vxy}\Gamma_{uvxy}L^{K}_{pr}L^{K}_{xy}v_{vr} = \sum_{v}T^{K}_{uv}S^{K}_{vp}.
\end{equation}

Regarding the last term, we notice that $\sum_{r}v_{vr}L^{K}_{ry}$ is the fully active part of the intermediate $\bm{S}^{K}$ of equation \ref{eq:int-s}. Hence, we define
\begin{equation}\label{eq:int-v}
 V^{K}_{ux} = \sum_{vy}\Gamma_{uxvy}S^{K}_{vy}
\end{equation}
and
\begin{equation}\label{eq:int-v-transp}
 \sum_{vy}\Gamma_{uxyv}S^{K}_{vy} = \sum_{vy}\Gamma_{xuvy}S^{K}_{vy} = (\bm{V}^{T})_{ux},
\end{equation}
where in equation \ref{eq:int-v-transp} we exploited the symmetry of the two-body reduced density matrix.
Gathering together equation \ref{eq:int-ts}, \ref{eq:int-v}, and \ref{eq:int-v-transp}, we can rewrite equation
\ref{eq:qt} as
\begin{equation}\label{eq:qt-cd-nice}
 \Qt_{up} = \sum_{r}Q_{ur}v_{pr}
            +\sum_{K}^{N_{\text{ch}}}\left[(\bm{T}^{K}\bm{S}^{K})_{up} + (\bm{X}^{K}\bm{L}^{K})_{up}\right]
\end{equation}
where $X^{K}_{ux} = V^{K}_{ux} + V^{K}_{xu}$. 
The evaluation of the transformed matrix requires thus $\mathcal{O}(N_{\rm ch}(N_{\rm act}^4 + N_b^2 N_{\rm act} + N_b N_{\rm act}^2))$ floating point operations, that can all be performed using optimized level 2 and 3 BLAS routines and $\mathcal{O}(N_b)^2$ words of memory for scratch.

\par
A remarkable fact about the CD is that all the operations involving different Cholesky vectors can be performed independently. As a consequence, parallelization is easily achieved by distributing the Cholesky vectors among the available processors, with a final reduction to be performed on the computed quantities. For this reason, we implemented the Cholesky loop as the most external in order to parallelize it with shared-memory (OpenMP) directives\cite{openmp}.

\section{Benchmarks}\label{sec:bench}
In this section we present benchmark calculations to illustrate the performance of the CD-CASSCF implementation. 
In all the calculation, convergence is achieved when the root mean square norm of both the orbital and CI gradient is below $10^{-7}$. The threshold for the Cholesky decomposition has been set to $10^{-4}$. The starting MOs are obtained from Restricted-Hartree-Fock (RHF) calculations. Point-group spatial symmetry was not used.
This section is organized as follows. In Section~\ref{sec:cfr}, we compare the CD and standard implementations on a small set of medium-sized molecules. In Section~\ref{sec:aromatic}, we present benchmark results on a set of medium-sized aromatic molecules, taken from ref. \citenum{Kreplin2019}, discussing in detail the cost associated with the various operations involved in the NEO CASSCF calculation. Finally, in Section~\ref{sec:larger}, we perform benchmark calculations on larger molecular systems, both in terms of number of basis functions and size of the active space.

\subsection{Cholesky versus standard implementation\label{sec:cfr}}
As a first analysis, we compare the performance of the Cholesky algorithm with respect to the 
standard one both in terms of computational cost and memory requirements. We selected 7 medium-sized molecular systems, catechol, naphthalene, pyridoxamine, 5,7‐dimethylidene‐2H,3H,5H,7H‐thieno[3,4‐b][1,4]dioxine (herein referred to as 2Me4HSdiox), indole, tryphtophan, and nicotine. The geometries were taken from ref. \citenum{Menezes2016}. 
The basis set used is Dunning's \textit{cc}-pVTZ\cite{Dunning1989}.
The standard CASSCF implementation handles the one- and two-electron integrals by reading them from disk at each macroiterations. As disk I/O is poorly parallelizable on standard hardware, it is a serial implementation, contrary to its CD-based counterpart. For consistency, we compare it to CD calculations run sequentially on a single CPU core. Both traditionals and CD CASSCF calculations were run on a single core of an Intel Xeon Gold 5120 CPU running at 2.20 GHz, on a cluster node equipped with 128GB of RAM. 
The standard implementation uses a semi-direct algorithm, where the MO Hessian is explicitly calculated. Given the size of the systems involved in this comparison, this is the most efficient algorithm. On the contrary, the CD implementation exploits a fully direct procedure as described in the previous section. For all calculations with both the standard and CD codes, the relevant MO-transformed ERIs and Cholesky vectors, respectively, were kept in memory.

The exact CD of the ERI matrix would generate $N_b(N_b+1)/2$ Cholesky vectors. We define a compression rate 
\begin{equation}
    \label{eq:compression}
    f = \frac{N_{b}(N_{b}+1)/2}{N_{\text{ch}}}
\end{equation}
to measure the effectiveness of the truncated CD in reducing the dimension of the computational problem.
We report in Table \ref{tab:cd-vs-std}, for each of the selected molecules, the active space (CAS), the number of basis functions ($N_{b}$), the disk space needed to store the two-electron integrals, the compression rate, the average time (averaged over the number of macroiterations) spent to perform the AO to MO integral transformation (AO to MO), the number of macroiterations, and the total CPU wall time (Time) in minutes.
\begin{table}[ht]
    \centering
    \begin{tabular}{lcccccccccc}
        \toprule
        & & & & \multicolumn{2}{c}{Size (GB)} & \multicolumn{2}{c}{AO to MO (min)} & & \multicolumn{2}{c}{Time (min)} \\
        \cmidrule(lr){5-6}\cmidrule(lr){7-8}\cmidrule(lr){10-11}
         molecule & CAS & $N_b$ & $f$ & CD & STD & CD & STD & It. & CD & STD  \\
         \midrule
catechol	& 10,8	& 324	& 24.88	& 0.9	& 18.3	& 0.12	& 4.03	& 10	& 10.0	& 46.1 \\
naphthalene	& 10,10	& 412	& 31.98	& 1.8	& 43.1	& 0.31	& 17.6	& 6	& 4.7	& 111.8 \\
nicotina	& 6,6	& 556	& 43.46	& 4.4	& 139.7	& 1.0	& 74.64	& 8	& 23.5	& 624.7 \\
triptofano	& 10,9	& 618	& 47.75	& 6.1	& 170.6	& 1.54	& 145.9	& 10	& 45.5	& 1520.2 \\
pyridoxamine	& 8,7	& 528	& 41.00	& 3.8	& 108.8	& 0.72	& 78.91	& 9	& 43.3	& 838.5 \\
2Me4HSdiox	& 12,9	& 446	& 34.10	& 2.3	& 61.8	& 0.43	& 27.35	& 10	& 27.0	& 300.4 \\
indole	& 10,9	& 368	& 28.37	& 1.3	& 30.4	& 0.2	& 13.24	& 6	& 3.7	& 82.5 \\
         \bottomrule
    \end{tabular}
    \caption{Systems used for the comparison of the standard and CD implementations of CASSCF in CFOUR. For each molecule, we report the size of the active space (CAS), the number of basis functions ($N_b$), the compression factor $f$ defined in eq. \ref{eq:compression}, the size in gigabytes of the Cholesky vectors (CD) and two electron integrals (STD), the average time in minutes required for the tranformation of the ERIs from the AO to the MO basis (AO to MO) for the CD and standard (STD) implementations, the total number of macroiterations, and the overall elapsed (wall) time in minutes for the calculation using the CD or standard (STD) code, respectively.}
    \label{tab:cd-vs-std}
\end{table}
As expected, the storage requirements for the CD vectors are significantly lower than for the standard two-electron integrals. It is worth remarking that even for the largest system of this set, the Cholesky vectors can easily be kept in memory even on a standard desktop computer. This is one of the main advantages of using a reduced order approximation of the ERIs, as it allows to perform full in-core calculations, avoiding thus slow disk I/O operations.
The traditional calculation is dominated in cost by the transformation of the integrals into the MO basis, which has to be repeated at each macroiteration. In the standard code, the AO to MO transformation is performed by reading a batch of integrals from disk and then transforming the first index, which is restricted to internal and active orbitals. The other three indices are then transformed using level 3 BLAS matrix-matrix multiplications (DGEMM), making the first step the cost-dominating one. The total cost of this operation is $\mathcal{O}(MN_b^4)$, where $M$ is the number of internal and active orbitals.

In the CD code, on the contrary, the Cholesky vectors are transformed in core by performing two DGEMM matrix-matrix multiplications per vector. No disk I/O is required and the overall transformation can be carried out in an efficient, highly vectorized fashion. The total cost of the CD AO to MO transformation is $\mathcal{O}(N_{\rm ch}N_b^3)$. It is apparent from Table \ref{tab:cd-vs-std} how this operation is no longer a bottleneck (see the next section for further discussions), thanks to both the compression stemming from the CD and to the handy data structure of the Cholesky vectors, that allows the procedure to be performed very efficiently. 
The reduction in the AO to MO timing is reflected also on the total execution time, which is one to two orders of magnitude lower in the CD code.

The chosen decomposition threshold ($10^{-4}$) allows us to obtain high compression rates while retaining an overall good accuracy. In Table \ref{tab:ene_cd_std}, we report the converged CASSCF energy obtained with both the Cholesky and standard implementations for the previously selected molecules. As it can be seen from the table, the two results are in agreement to at least the fourth decimal digit, with the largest deviation being about $50\mu E_h$. We note that it has been documented in the literature that CD benefits from error cancellation, thus further increasing the accuracy of energy differences\cite{Aquilante2008,Aquilante2007}.
\begin{table}[ht]
    \centering
    \begin{tabular}{lcccc}
    \toprule
    molecule & CD Energy & STD Energy \\
    \midrule
    catechol	& -380.625~850~38	& -380.625~855~05\\
    naphthalene	& -383.592~108~03	& -383.592~112~05\\
    nicotine	& -495.948~509~96	& -495.948~509~88\\
    tryptophan	& -682.506~309~98	& -682.506~290~53\\
    pyridoxamine& -568.806~659~19	& -568.806~674~05\\
    2Me4HSdiox	& -855.037~280~36	& -855.037~328~27\\
    indole		& -361.678~956~49	& -361.678~962~79\\
    \bottomrule
    \end{tabular}
    \caption{Comparison between the converged CASSCF energy of the Cholesky and standard implementation. Energy values are in given in Hartree. }
    \label{tab:ene_cd_std}
\end{table}

\subsection{Benchmark calculations for medium-sized systems\label{sec:aromatic}}
The first benchmark set is composed of 21 aromatic molecules. The geometries were taken from Ref. \citenum{Menezes2016};
the set was used also by \citeauthor{Kreplin2019} to test their MCSCF solver \cite{Kreplin2019,Kreplin2020}. 
As before, the basis set used is Dunning's \textit{cc}-pVTZ\cite{Dunning1989}, for a total of 250 basis functions for the smallest molecule and 618 for the largest one. 
For each system, all the orbitals, including the core ones, are fully variationally optimized. All the calculations presented here were performed on a single cluster node equipped with two Xeon Gold 5120 CPUs, for a total of 28 cores, running at 2.20GHz. Shared memory parallelization is exploited in all the calculations. We point out here that we do not expect the implementation to be fully scalable, the limiting factor being the Full CI code. This is due to the fact that the sequential code is highly cache-optimized, which causes an overload of the cache, and consequent loss of efficiency, when more cores of the same processor share cache access. Nevertheless, even a simple-minded OpenMP parallelization of the main loops is beneficial. 
In Table \ref{tab:set-1} we reported for each molecule, the active space (CAS), the number of basis functions, the number of macro-iterations required to converge, and the total CPU wall time in minutes. 
\begin{table}[ht]
    \centering
    \begin{tabular}{lcccc}
        \toprule
        molecule       & CAS   &  $N_{b}$   &  It.    &  Time (min) \\ 
        \midrule
        adrenaline	& 10,8	& 572	& 11	& 19.4 \\
        azulene	& 10,10	& 412	& 6	& 1.6 \\
        biphenyl	& 12,12	& 500	& 10	& 5.3 \\
        catechol	& 10,8	& 324	& 10	& 3.1 \\
        dopamine	& 10,8	& 484	& 10	& 13.7 \\
        indole	& 10,9	& 368	& 6	& 1.0 \\
        l-dopamine	& 10,8	& 574	& 9	& 19.6 \\
        naphthalene	& 10,10	& 412	& 6	& 1.3 \\
        niacin	& 8,8	& 340	& 10	& 1.7 \\
        niacinamide	& 8,8	& 354	& 11	& 3.6 \\
        nicotine	& 6,6	& 556	& 8	& 7.0 \\
        nor-adrenaline	& 10,8	& 514	& 11	& 15.5 \\
        picolinic acid	& 8,8	& 340	& 7	& 1.2 \\
        pyridine	& 6,6	& 250	& 6	& 0.2 \\
        pyridoxal	& 8,8	& 486	& 8	& 4.9 \\
        pyridoxamine	& 8,7	& 528	& 9	& 13.2 \\
        pyridoxin	& 8,7	& 514	& 9	& 13.3 \\
        serotonin	& 12,10	& 558	& 11	& 20.3 \\
        tryptophan	& 10,9	& 618	& 10	& 11.2 \\
        2Me2HSdiox	& 10,7	& 474	& 9	& 12.1 \\
        2Me4HSdiox	& 12,9	& 446	& 10	& 8.3 \\ 
        \bottomrule
    \end{tabular}
    \caption{First aromatic benchmarks set results. For each molecule, the active space (CAS), the number of basis functions ($N_{b}$), the number of macro-iterations (It.), and the total CPU wall time (Time) in minutes are presented. 2Me2HSdiox is the abbreviation for 5,7‐dimethyl‐2H,3H‐thieno[3,4‐b][1,4]dioxine.}
    \label{tab:set-1}
\end{table}
As a first consideration, we note that the starting RHF orbitals used in this benchmark are a poor choice for CASSCF, as they cause the MO Hessian to be poorly conditioned. Using RHF orbitals is therefore a good way to test the robustness of the optimization algorithm, but not an optimal one for application sake. In practice, the ill-conditioning of the Hessian reflects both into slow convergence of the microiteration, which are Davidson iterations used to compute the lowest eigenvector of the NEO augmented Hessian, and in a larger number of macroiterations. Despite such difficulties, all calculations converged in at most $11$ iterations and took less than 20 minutes, which demonstrates the robustness of the NEO algorithm and the overall efficiency of the implementation.

To further investigate the performance of the algorithm, we can subdivide the work into three main tasks---the AO to MO transformation, the optimization of the MOs (MOs opt.), and the optimization of the CI coefficients (CI opt.). The MOs optimization includes the calculation of the orbital gradient (eq. \ref{eq:orb-grad}), which in turn requires to assemble the various Fock matrices (eqs. \ref{eq:fi} and \ref{eq:fa}-\ref{eq:q}), the calculation of the diagonal of the MO Hessian (which is used as the preconditioner in the Davidson diagonalization), and the evaluation of the direct equations \ref{eq:co-block} and \ref{eq:oo-block} for each micro-iterations. On the other hand, the CI optimization consists in computing the reduced density matrices, assembling the CI gradient (eq. \ref{eq:ci-grad}), and evaluating equations \ref{eq:oc-block} and \ref{eq:cc-block} at each micro-iterations. Table \ref{tab:set-1-spec} shows the percentage time to perform these three operations with respect to the total time of a specific macro-iteration. Also, the total number of iterations required to solve the NEO problem (micro-It.) is reported.  
\begin{table}[ht]
    \centering
    \begin{tabular}{lccccc}
        \toprule
        molecule & AO to MO & MOs opt. & CI opt. & micro-It. & Time (s)\\
        \midrule
        catechol	& 3.83	& 95.45	& 0.41	& 19	& 13.8 \\
        naphthalene	& 5.25	& 90.14	& 4.22	& 19	& 25.3 \\
        nicotine	& 6.09	& 93.78	& 0.01	& 16	& 71.7 \\
        tryptophan	& 5.74	& 93.88	& 0.25	& 19	& 120.5 \\
        biphenyl	& 3.16	& 66.88	& 29.47	& 16	& 95.8 \\
        \bottomrule
    \end{tabular}
    \caption{Percentage time of the three leading operations with respect to the total time of a specific macroiteration (Time). AO to MO refers to the atomic orbitals to molecular orbitals transformation of the Cholesky vectors, MOs opt. is the time spent in the MOs optimization and includes operations such as: calculation of the orbital gradient, evaluation of the NEO augmented Hessian-orbital trial vector products. CI opt. refers to the CI optimization and include the following operations: calculation of the CI gradient, calculation of the reduced density matrices, and evaluation of the NEO augmented Hessian-configuration trial vector products. micro-It is the number of microiterations required to solve the NEO eigenvalue-eigenvector problem.}
    \label{tab:set-1-spec}
\end{table}
The CD extremely facilitate the integrals transformation shifting the bottleneck to the MOs optimization part. For the systems considered, in particular, most of the time is spent in computing the transformed Fock matrices, an operation that is required to assemble the NEO Hessian-orbital trial vector product (eqs. \ref{eq:co-block}, \ref{eq:oo-block}). We also note that for larger active spaces, such as in biphenyl, the cost associated with the CI part starts to become non negligible. 

\subsection{Benchmark calculations for large systems}\label{sec:larger}
\begin{figure}[ht]
    \centering
    \includegraphics[width=.7\textwidth]{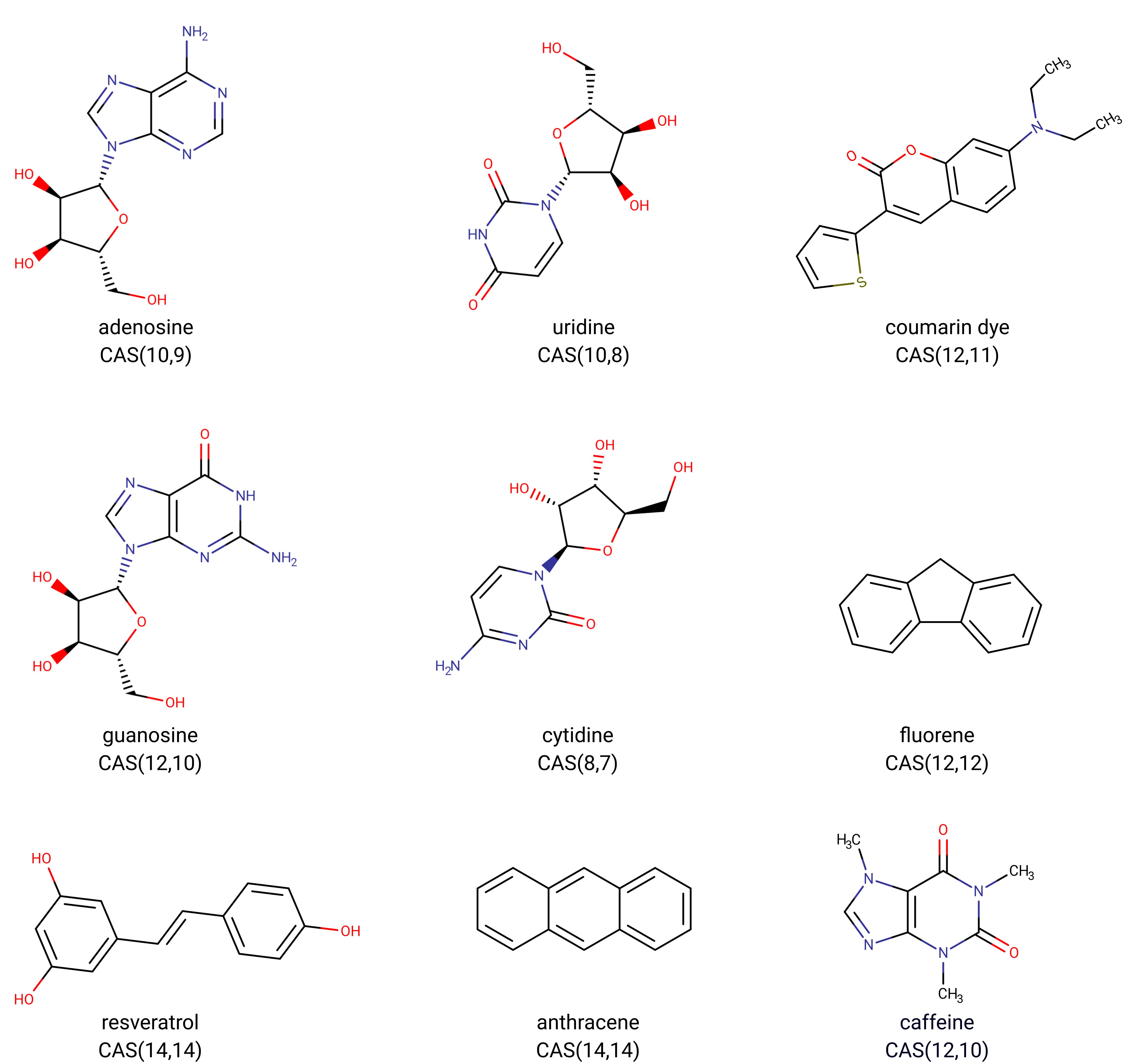}
    \caption{New set of aromatic molecules used to test the CD-CASSCF algorithm together with their active spaces.}
    \label{fig:new-set}
\end{figure}

\begin{figure}[ht]
    \centering
    \includegraphics[width=.5\textwidth]{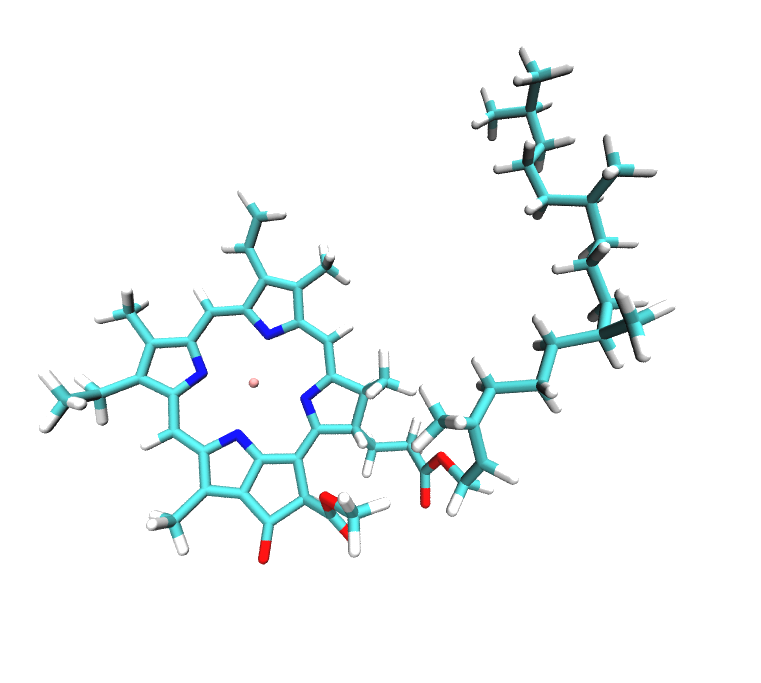}
    \caption{Structure of the chlorophyll molecule.}
    \label{fig:chl}
\end{figure}

In order to test the new implementation on more challenging problems, we augment the benchmark set discussed in sec. \ref{sec:aromatic} with 10 larger molecules. The calculations involved active spaces up to CAS(14,14) and as many as 2962 basis functions. The geometries of the molecules were optimized at the B3LYP/6-31G(d)\cite{Becke1993,Hehre1972} level of theory using the Gaussian 16 suite of programs\cite{g16}. All the structures can be found in the Supporting Information, a pictorial representation of the molecules is given in Figures \ref{fig:new-set}, and \ref{fig:chl}. For all CD-CASSCF calculations, we used Dunning's cc-pVTZ basis set. The calculations were performed on the same cluster node used for the previous set, with the exception of the largest system (chlorophyll), for which we used a cluster node equipped with 1.2TB of memory and 4 Intel Xeon Gold 6140M CPUs running at 2.30GHz, for a total of 72 cores.

\begin{table}[ht]
    \centering
    \begin{tabular}{lcccc}
        \toprule
        molecule       & CAS   &  $N_{b}$   &  It.    &  Time (m)\\ 
        \midrule
        adenosine	& 10,9	& 752	& 8	& 17.0 \\
        caffeine	& 12,10	& 560	& 7	& 6.8 \\
        coumarin dye	& 12,11	& 872	& 12	& 48.0 \\
        cytidine	& 8,7	& 692	& 8	& 15.8 \\
        fluorene	& 12,12	& 530	& 7	& 5.0 \\
        guanosine	& 12,10	& 782	& 9	& 20.7 \\
        uridine	& 10,8	& 678	& 8	& 17.4 \\
        anthracene	& 14,14	& 560	& 13	& 38.8 \\
        resveratrol	& 14,14	& 678	& 12	& 67.3 \\
        chlorophyll & 10,10 & 2962 & 15 & 917.7\\
        \bottomrule
    \end{tabular}
    \caption{Results for the new set of aromatic molecules. For each molecule we reported the active space (CAS), the number of basis functions ($N_{b}$), the number of macroiterations (It.), and the total elapsed CPU (wall) time in minutes.}
    \label{tab:new-set}
\end{table}
The total execution times for the first seven systems are comparable with the ones obtained for the previous set and show the overall good performance of the code, both in terms of total time and of convergence. The calculations on anthracene and resveratrol, for which a large CAS(14,14) active space (that consists of 11~778~624 Slater determinants) was used, are dominated by the cost of the CI-related operations. In order to show this, we report in Table \ref{tab:large-cas-spec} the percentage of the time spent performing the same operations discussed in Section~\ref{sec:aromatic} for the two molecules for the last macroiteration. 
\begin{table}[ht]
    \centering
    \begin{tabular}{lccccc}
        \toprule
        molecule & AO to MO & MOs opt. & CI opt. & micro-It. & Time (min)\\
        \midrule
        anthracene	& 0.81	& 14.44	& 83.5	& 11	& 9.2 \\
        resveratrol	& 1.15	& 15.31	& 82.59	& 13	& 16.4 \\
        \bottomrule
    \end{tabular}
    \caption{Percentage time of the three main operations (AO to MO, MOs opt., CI opt.) with respect to the total time of the last macro-iteration. }
    \label{tab:large-cas-spec}
\end{table}
Here, the most expensive operations are the direct-CI steps needed to compute the CI gradient and the CI part of the NEO augmented Hessian-configuration trial vector products, together with the assembling of the reduced density matrices. As it can be seen, these operations take about 80\% of the total time.

Finally, the largest system tested, chlorophyll, can be thought as a pilot example of a large-scale application on a biologically relevant molecule. For this specific case, the storage of the Cholesky vectors in memory required more than 600GB of RAM, which is the reason why the calculation was performed on a different computer. The calculation converged in 15 macroiterations and took slightly more than 15 hours. It is interesting to look in more detail the cost associated to the various operations in a given macroiteration. Focusing on the ninth macroiteration as an example, which required 12 microiterations to converge, the AO to MO transformation of the Cholesky vectors took 11.7 minutes, and the MOs optimization lasted 54.9 minutes. The time required by the CI operations is negligible. Here, we see clearly that the price to pay for a second-order optimization lies in the cost of solving the NEO equations, which are by far dominating the overall cost of the calculation. Nevertheless, the guarantee of convergence remains an attractive feature of the method and the overall time required for the calculation is not excessive. 



\section{Conclusions}\label{sec:concl}
We have presented the implementation of a second-order CASSCF optimization algorithm that exploits the Cholesky Decomposition of the two electron integrals. The algorithm is based on a trust-region method, which requires to solve diagonally shifted Newton-Raphson equations known as Levenberg-Marquard (LM) equations. Also, it adaptively modifies the trust radius during the optimization according to the value of the energy with the result that the overall algorithm always converges to the closest minimum for regular enough functions. The coupling between orbitals and CI coefficients is naturally included in the off-diagonal blocks of the Hessian matrix making this algorithm naturally second-order in all parameters. The implementation is based on the Norm-Extended Optimization (NEO) formalism where the LM equations are recast into an eigenvalue problem, where the first eigenvector provides the optimal direction for ground-state minimization problems.  

To reduce the computational cost associated with orbitals optimization, which is dominating for not-too-large active spaces, we implemented the NEO algorithm using the Cholesky Decomposition (CD) of the two-electron integrals matrix. The NEO equations were rewritten in terms of the Cholesky vectors, taking particular care in recasting all the equations in a way that allowed us to implement them efficiently thanks to an extensive use of level 3 BLAS routines.
The implementation exploits a fully direct algorithm where the Hessian matrix is never explicitly calculated. Furthermore, since the Cholesky vectors are independent among the others, the code has been parallelized with shared-memory OpenMP directives.  

The resulting algorithm was tested on various aromatic systems. We used a triple zeta basis set with up to 2962 functions and active spaces up to CAS(14,14). Despite the choice of a very poor guess for the orbitals, namely, Restricted-Hartree-Fock canonical orbitals, all the calculations converged swiftly and required limited computer time. Thanks to the effective compression of the two-electron integrals matrix operated by the CD, fully in-core calculations are possible for most systems, eliminating thus the bottleneck of slow disk I/O. While several further improvements and optimizations are possible, for instance, to improve the convergence of the microiterations, the benchmark calculations reported in this contribution show that a rigorous second-order algorithm can be used in large-scale applications at a reasonable computational cost. 
Future work will focus on both algorithmic improvements and extensions of the methodology. In particular, a first order procedure such as super CI\cite{Roos1980,Siegbahn1981} could be used in the preliminary phase of a calculation to achieve an initial intermediate convergence goal, thus providing a very good starting point for the quadratically convergent optimization. We also plan to extend the second order procedure to the simultaneous optimization of several electronic states and to the calculation of analytical gradients, by implementing differentiated Cholesky vectors\cite{Delcey2014,Feng2019}. 

\begin{acknowledgement}
T.N. acknowledges the traineeship funds from the Erasmus+ program. 
\end{acknowledgement}

\begin{suppinfo}

Explicit equations for the direct product with the MO Hessian, implementable expressions for the transformed Fock matrices, and optimized molecular geometries for the new set of aromatic molecules used as benchmarks.

\end{suppinfo}

\bibliography{cdcas-energy}

\end{document}